\documentstyle[epsfig,11pt]{article}

\input epsf
\title{\bf On the inclusive gluon jet production from the triple pomeron
vertex in the
perturbative QCD}
\author{M.A.Braun\\
Dep. of High Energy physics,
 University of S.Petersburg,\\
198904 S.Petersburg, Russia}
\def\beq{\begin{equation}}
\def\eeq{\end{equation}}
\def\noi{\noindent}
\def\fa{f_1(r_1)}
\def\fb{f_1(r_2)}
\def\fc{f_2^*(r_1)}
\def\fd{f_2^*(r_2)}
\def\ha{h_1(r_1)}
\def\hb{h_1(r_2)}
\def\hc{h_2^*(r_1)}
\def\hd{h_2^*(r_2)}
\def\ka{k_1(r_1)}

\def\pa{X}
\def\pb{Y}
\def\pc{U}
\def\pd{V}
\def\pz{Z}

\def\rr{r_1\leftrightarrow r_2}

\begin{document}
\maketitle
\medskip
\noi{\bf Abstract.}
Single and double inclusive cross-sections for gluon jet production
from within the triple pomeron vertex are studied in the reggeized
gluon technique. It is shown that to satisfy the AGK rules the
vertex has to be fully symmetric in all four reggeized gluons
which form the two final pomerons. The single inclusive cross-sections are
found for different cuttings of the triple pomeron vertex. They sum
into the expression obtained by Yu.Kovchegov and K.Tuchin in the colour
dipole picture. The found double inclusive cross-sections satisfy the AGK
rules.

\vspace{0.5cm}

\section{Introduction}
In the perturbative QCD at small values of $x$
the strong interaction  can  be modelled by
the exchange of reggeized gluons and BFKL pomerons as their bound states.
In the limit  of large number of colours $N_c\to\infty$ the model reduces
to the propagation and triple interaction of pomerons in the tree diagram
approximation. The equations which sum these diagrams have been written
both for DIS (BK equation ~\cite{bal,kovch,bra1}) and for nucleus-nucleus
collisions ~\cite{bra2}. The solution of these equations allows to find
total cross-sections for processes like $\gamma^*A$ and $AB$.
The next important observables, which carry much more information about
the dynamics, are  inclusive cross-sections to produce
gluon jets which are to hadronize into the observed hadrons.
First calculations of single and double inclusive  cross-sections were
made in ~\cite{bra3} on the basis of the AGK rules \cite{AGK}.  From them it
follows in particular that in the single inclusive cross-section the
produced gluon jet  comes from within
the initial BFKL pomeron before its branchings.
Later from the dipole picture a slightly different expression
for the same cross-section was derived. In it, apart from the
above-mentioned naive AGK contribution, another term appeared,
which could be interpreted as emission from within the triple pomeron
vertex itself ~\cite{KT}. Such a contribution is not prohibited by the AGK
rules but was usually neglected as small. However in the perturbative QCD
at small $x$ it is of the same order as the emission from the pomeron.
Further analysis performed in \cite{bra4} in the framework
of reggeized gluon diagrams  seemingly discovered many terms in the
contribution to production from the vertex (and among them also the one found in
~\cite{KT} ). However the derivation in ~\cite{bra4} was based on certain
{\it  at hoc} assumptions, so that it was
stressed there that the derivation was in fact quite heuristic
and needed a more detailed study. This paper, which is a direct
continuation of ~\cite{bra4}, presents results of this study.

We find that a  more careful analysis of reggeized gluon diagrams and
especially the validity of the AGK rules for different forms in which their
sums may be presented leads to results which differ from those obtained in
~\cite{bra4}. We find that the triple Pomeron contribution in the form used
in that paper (with the so called diffractive vertex $Z$)
does not satisfy the AGK rules and only the form  with the symmetric Bartels
vertex $V$ does satisfy them \footnote {We highly
appreciate our numerous discussions
with J.Bartels who has always insisted on this point (see also \cite{BR})}
This circumstance radically changes the  derivation of the
contribution to jet production from within the triple pomeron vertex.
The found terms  for different cuttings of the pomerons joined by the
vertex are as numerous and complicated as in ~\cite{bra4}
but in the sum they indeed combine into the Kovchegov-Tuchin term.
Thus the reggeized gluon digrams approach to the gluon jet
production completely agrees with the dipole picture. In fact the most
convenient technique combines both approaches and allows to obtain results
in the simplest way. Using it one is able to easily
construct an evolution equation for the 4-gluon amplitude with  jet
production. One can also
show that contributions
which blatantly violate the AGK rules found for the double inclusive
cross-section in ~\cite{JK} are in fact absent.

\section{The AGK rules}
We start with the scattering amplitude of some projectile
(e.g $\gamma^*$) on two scattering centers, which corresponds to the
triple interaction of BFKL pomerons and is schematically shown in Fig. 1,
where one can also see our notations for the initial and final momenta
$p_i$ and $p'_i$, $i=0,1,2$.
\begin{figure}
\hspace*{3 cm}
\epsfig{file=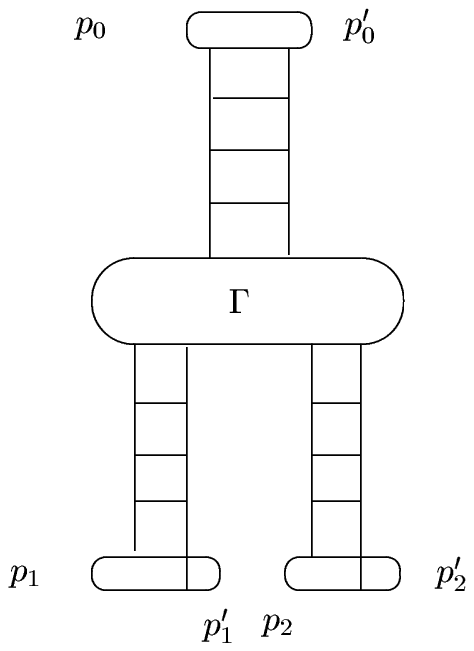,bbllx=208pt,bblly=527pt,bburx=415,bbury=716pt,width=5cm}
\caption{Triple pomeron contribution to the $3\to 3$ amplitude}
\label{fig1}
\end{figure}
Both the projectile and two
targets,
are supposed to be colorless objects like quark-antiquark
loops (photons or onia). In practice this amplitude, multiplied by
one half of the square of the nuclear profile function $T(b)$,  represents
the contribution from the double rescattering in the nucleus. However for
the problem at hand  this circumstance is unimportant.
We assume that the c.m, energy
$s=(p_0+p_1)^2=(p_0+p_2)^2$ is large and the transferred momenta
$t_i=(p'_i-p_i)^2$ are finite and so much smaller than $s$. In
fact in the following we shall concentrate on the forward case
$t_0=0$ which is of most practical importance.
The  ladders represent the initial and two final BFKL pomerons
and the central blob corresponds to the triple pomeron vertex, which is
local in rapidity.
 Allowing for the pomerons to be Regge cuts and
not simple poles (as is the case of the BFKL pomeron)
and for the two lower pomerons to have different
energies one gets a representation for the amplitude ~\cite{BW}
\beq
T_{3\to 3}=\frac{1}{2\pi^4}
\frac{s_1s_2}{M^2}\int\prod_{k=0}^2\Big(\frac{dj_k}{2\pi i}
\zeta_k\Big)\left(\frac{s_1}{M^2}\right)^{j_1-1}
\left(\frac{s_2}{M^2}\right)^{j_2-1}\left(\frac{M^2}{Q^2}\right)^{j-1}
F(j_i,t_i).
\eeq
The signature factors are defined as
\beq
\zeta=-\pi\frac{e^{-i\pi j}+1}{\sin(\pi j)},
\eeq
where for $\zeta_0$ one should take $j=j_0-j_1-j_2$.
As compared to ~\cite{BW} we have included a factor $i$ for each
pomeron
since in the skeleton diagrams one should use elemental scattering
amplitudes multiplied by $i$.
For physical scattering we are to take $s_1=s_2=s$.
Function $F(j_i,t_i)$ corresponds to the diagram of Fig. 1 in the
complex angular momentum representation. It is a real function,
which is a  product of three pomerons
in the $j$-representation and the triple pomeron vertex $\Gamma$:
\[
F(j_i,t_i)=\sum_{a_i,b_i}\int \frac{d^2q_1}{(2\pi)^2}
\frac{d^2k_1}{(2\pi)^2}\frac{d^2k_3}{(2\pi)^2}\]\beq
\Gamma_{a_1a_2a_3a_4}^{b_1b_2}(k_1,k_2,k_3,k_4|q_1,q_2)
P_1^{a_1a_2}(j_1;k_1,k_2)P_2^{a_3a_4}(j_2;k_3,k_4)P_{b_1b_2}(j_0;q_1,q_2).
\eeq
Here $P$, $P_1$ and $P_2$ are the initial and two final pomerons,
$a_i$ and $b_i$ are their colour
indexes, $k_i$ and
$q_i$ are their transverse momenta, with $q_1+q_2=p_0-p'_0$,
$k_1+k_2=p_1-p'_1$ and $k_3+k_4=p_2-p'_2$.
Since the pomerons are colorless, $P$'s include a projector onto
the colourless state
\beq
P_{b_1b_2}(j_0;q_1,q_2)=\frac{1}{N_c^2-1}
\delta_{b_1b_2}P(j_0;q_1,q_2)
\eeq
and similar for the final pomerons. The normalization is chosen to include an
extra factor $\sqrt{N_c^2-1}$ into the function $P(q_1,p_2)$. This
normalization depends on the one chosen for the external sources. In the
following, to generalize for multiple scattering and relate the
pomerons to the sum of fan diagrams   $\Phi$, we rescale the final pomerons as
$P\to g^2P$ and the initial pomeron as $P\to P/g^2$.
Then for single scattering $\Phi=\Phi^{(1)}=PT(b)$ and the vertex $\Gamma$ in (3)
coincides with the  vertex in the BK equation.

We first demonstrate that the relation between the imaginary parts
of the amplitude coming from different cuts trivially follows
from the representation (1)
with a real function $F(j_i,t_i)$.
Note that
$
{\rm Im}\,\zeta=i\pi
$
and that for pomerons  in the lowest order
$
\zeta=i\pi.
$
Armed with these properties we may calculate the total and
partial imaginary parts of the amplitude. Since all non-trivial
dependence on energies is contained in signature factors, we have only
to follow their change when taking the relevant discontinuities.

Fist the total imaginary part. To find it we have just to
substitute all $\zeta$'s by $i\pi$,
which will result in a contribution
\beq
({\rm Im}\,T)^{tot}=-\frac{1}{2\pi}
\frac{s_1s_2}{M^2}\int\prod_{k=0}^2\frac{dj_k}{2\pi i}
\left(\frac{s_1}{M^2}\right)^{j_1-1}
\left(\frac{s_2}{M^2}\right)^{j_2-1}\left(\frac{M^2}{Q^2}\right)^{j-1}
F(j_i,t_i).
\eeq
Partial imaginary parts corresponding to different cuts.
The diffractive cut corresponds to taking $i\zeta_0=2\pi$ (just the
discontinuity divided by $i$), $\xi_1=\xi_2=-\pi$ and dividing the whole
expression by two. Obviously we get
\beq
({\rm Im}\,T)^{dif}=-({\rm Im}\,T)^{tot}.
\eeq
The double cut corresponds to taking  $i\zeta_0=i\zeta_1=i\zeta_2=2\pi$
(again the discontinuities divided by $i$) and dividing the whole
expression by 2$\times$ 2 (2 for the imaginary part and 2 for the
identity of the two legs). Obviously we get twice the diffractive
cut
\beq
({\rm Im}\,T)^{double}=2\,({\rm Im}\,T)^{dif}.
\eeq
Finally single cuts correspond to $\xi_0$ and one of $\xi_{1,2}$
substituted by $2\pi$ the other one by $-\pi$ multiplying by
2 for the complex conjugate part and dividing by 2 to pass from
the discontinuity to the imaginary part. As a result we get four
times the diffractive cut with the opposite signs
\beq
({\rm Im}\,T)^{single}=-4\,({\rm Im}\,T)^{dif}.
\eeq
The sum of these partial contributions is equal to the total
imaginary part ant their relative weights  correspond to the AGK rules.

Note however that what we have just presented is only a formal
derivation of the AGK rules. In fact one has to be able to identify
intermediate states and production amplitudes
which generate different cuts of the amplitude in the unitarity relation
for the total ${\rm Im}\,T$. This is trivial for the
internal gluons in the pomerons themselves, but not so for the coupling
of the pomerons to the external particles and to each other. Such an
identification is not needed when one studies the total amplitude and the
cross-section which it describes, but becomes a necessity if one wants
to see which particles are produced from the vertexes describing these
couplings.
The problem mostly concerns  the double cut in variables
$s_1$ and $s_2$ in (1), which has to be reinterpreted as a single
unitarity cut.

That this is not generally possible illustrates the diagram shown in Fig.
2, say for a scalar theory with a triple interaction. It obviously
contains the double cut in lower reggeons energies, but this cut  cannot
be identified with the unitarity cut, which cannot pass through both
reggeons.  So we have to study the unitarity relation for the amplitude
in correspondence with the cuts we have discussed.
\begin{figure}
\hspace*{6cm}
\epsfig{file=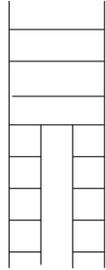,width=1.3cm}
\caption{Amplitude which possesses a double cut which is not unitary}
\label{fig2}
\end{figure}
Let us start from the diffractive cut, which  is graphically shown in
Fig. 3$a$,
where also the corresponding
intermediate states in the unitarity relation for the three participating
pomerons are shown. The cut drawn  through the three-pomeron vertex
is purely symbolical: it is not possible to interprete the cut vertex
$\Gamma$
terms of intermediate states for the production amplitudes in a
straightforward manner. The general
formula (1) only tells us that the cut vertex is a real function
independent of the cutting
$\Gamma_{a_1a_2a_3a_4}^{b_1b_2}(k_1,k_2,k_3,k_4|q_1,q_2)\equiv
\Gamma(1,2,3,4)$, where the numbers refer to the final gluons, 1 and 2
in the first lower pomeron and 3 and 4 in the other one.
%
\begin{figure}
\hspace*{3 cm}
\epsfig{file=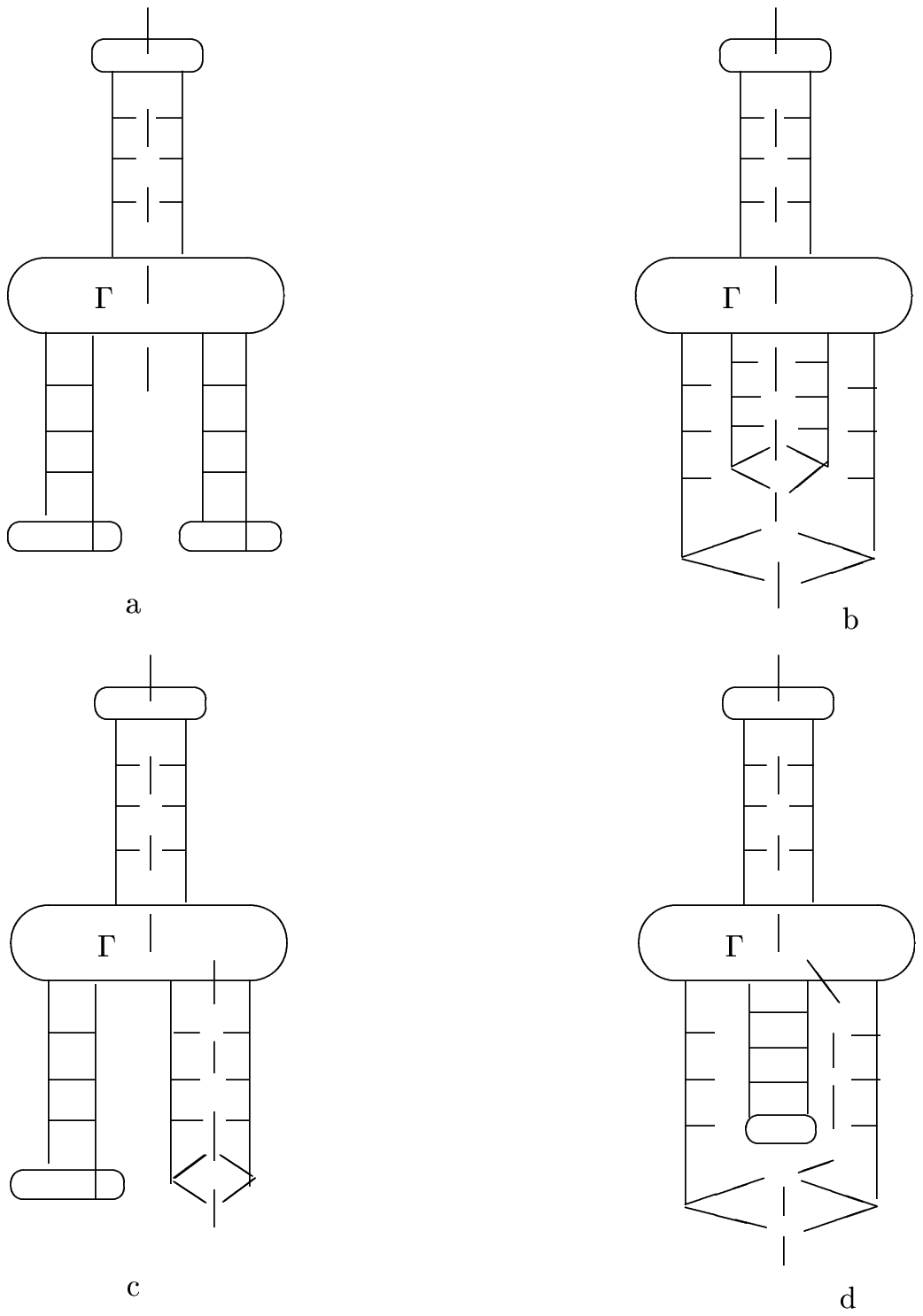,bbllx=151pt,bblly=235pt,bburx=491,bbury=694pt,width=8cm}
\vspace*{-1cm}
\caption{Diffractive ($a$), double ($b$) and single ($c$ and $d$) cuts
and their unitarity content}
\label{fig3}
\end{figure}
The double cut is shown in Fig. 3$b$  together with its
unitarity content. Again the cut through the vertex  is not directly
expressible in terms of intermediate states. One observes that it
corresponds to the interchange of the final gluons 2 and 3 in the vertex:
$\Gamma(1,3,2,4)$. However, as
mentioned, the cut vertex should not depend on the particular cutting.
It should also be symmetric in pairs of gluons (1,2) and (3,4) due to
the properties of the pomeron. As a result,
the vertex function $\Gamma$
has to be completely symmetric in all four final gluons 1,2 3 and 4.
The single cut contributions shown in Fig.3$c$ and $d$ do not imply any
new condition
on the vertex function $\Gamma$
So we conclude  that  the necessary condition for the fulfillment
of the AGK rules  for the triple pomeron contribution represented
according to Eq. (1) is the complete symmetry of the triple pomeron vertex
in all four final  reggeized gluons. This requirement generalizes the one
in the original AGK derivation that the vertex should not change
with different cuttings.

Note that the contributions to the total imaginary part of the amplitude
$T_{3\to 3}$ can be classified not only by a particular cutting,
diffractive, double  or single, but also by
the number of gluons emitted at the vertex rapidity. In the lowest order
in can be zero or unity.  So each of the contributions from a
particular cutting
can be split in two
\beq
({\rm Im}\,T)^{cut}=\sum_{n=0,1}({\rm Im}\,T)^{cut}_n,
\eeq
where $cut\,=\, diffractive,\, double,\, single$ and $n$ is the number of gluons
emitted at the vetex. It is important that although the sums (9)
satisfy the simple AGK relations (6)-(8), the separate contributions from
$n=0$ and $n=1$  do not, as we shall see in the following.

\section{The triple pomeron contribution in the perturbative QCD}
Analysis of reggeized gluon diagrams shows that the amplitude
with four final reggeized gluons may be represented in different forms.
From the direct study of the triple discontinuity with two, three and
four exchanged reggeized gluons one finds an expression which is a sum
of the double pomeron exchange and  triple pomeron contribution with the
so-called
diffractive vertex $Z$ \cite{BW,bra5,BBV}. However neither the
vertex $Z$ nor the gluon coupling to the external particles in the double
pomeron exchange contribution are  symmetric
in all four reggeized gluons 1,..,4 (they are only symmetric in pairs 12
and 34). So according to the results of the previous section
neither of these two contributions can separately satisfy the AGK rules.
However these two
contributions can be transformed
in another two, one of which  (the 'reducible' part) has a form of a single
pomeron exchange
and the other (the irreducible' part) of a triple pomeron contribution with a
different vertex $V$
symmetric in all the gluons \cite{BW}. Remarkably in the high-colour limit
this vertex coincides with the one introduced by A.Mueller and B.Patel
in the colour
dipole model \cite{MP} and which represents the pomeron interaction in the
hA and AB collisions. Our results show that it is this vertex for
the triple pomeron interaction which satisfies the AGK rules.
Note that this statement was made rather long ago in connection with the
amplitude for the scattering on a single center \cite{BR,BW}. Our
results  generalize it to the scattering on two (or many) centers.

For our purpose we only need the vertex projected onto the incoming
pomeron colourless color state
\beq
\sum_b\Gamma_{a_1a_2a_3a_4}^{bb}(k_1,k_2,k_3,k_4|q_1,q_2)
\equiv \Gamma_{a_1a_2a_3a_4}(k_1,k_2,k_3,k_4|q_1,q_2).
\eeq
This vertex is related
to the  symmetric Bartels vertex $V$ by the relation
\beq
g^2\Gamma_{a_1a_2a_3a_4}(k_1,k_2,k_3,k_4|q_1,q_2)=\frac{1}{2}
q_1^2q_2^2V_{a_1a_2a_3a_4}(k_1,k_2,k_3,k_4|q_1,q_2).
\eeq
The vertex $V$ has the following colour and momentum structure
\[
V_{a_1a_2a_3a_4}(k_1,k_2,k_3,k_4|q_1,q_2)\]\beq=
\delta_{a_1a_2}\delta_{a_3a_4}V(1,2,3,4)+
\delta_{a_1a_3}\delta_{a_2a_4}V(1,3,2,4)+
\delta_{a_1a_4}\delta_{a_3a_2}V(1,4,3,2).
\eeq
For brevity we suppress the dependence on the momenta of the
initial pomeron $q_1,q_2$ common to all the terms and denote the
momenta of the final  gluons by their numbers,
so 1 means $k_1$ and so on. Function $V(1,2,3,4)$ is symmetric
under the interchanges 1$\leftrightarrow$2, 3$\leftrightarrow$4
and 12$\leftrightarrow$34. The whole expression (14) is obviously
completely
symmetric in all four gluons.

The explicit expression for the function $V(1,2,3,4)$ is conveniently given
in terms of function $G(k_1,k_2,k_3)$  introduced in \cite{BW} and
generalized to the non-forward direction in \cite{BV}:
\[
V(1,2,3,4)=\frac{g^2}{2}\Big(G(1,23,4)+G(2,13,4)+G(1,24,3)+G(2,14,3)-
G(12,3,4)\]\beq-G(12,4,3)-G(1,2,34)-G(2,1,34)+G(12,0,34)\Big),
\eeq
where again for brevity we denote 12=1+2=$k_1+k_2$ etc.
Function $G$ has the form
\[
G(k_1,k_2,k_3|q_1,q_2)=-g^2N_cK(k_1,k_2,k_3|q_1,g_2)
\]\beq-(2\pi)^3\delta^2(q_1-k_1)\Big(\omega(2)-\omega(23)\Big)
-(2\pi)^3\delta^2(q_2-k_3)\Big(\omega(2)-\omega(12)\Big).
\eeq
Here $\omega(k)$ is the gluon Regge trajectory
and the kernel for transition of two gluons into three $K$
is given by
\beq
K(k_1,k_2,k_3|q_1,q_3)=\frac{(k_1+k_2+k_3)^2}{q_1^2q_3^2}
+\frac{k_2^2}{(k_1-q_1)^2(k_3-q_3)^2}-\frac{(k_1+k_2)^2}{q_1^2(k_3-q_3)^2}
-\frac{(k_2+k_3)^2}{q_3^2(k_1-q_1)^2}.
\eeq
It conserves the momentum, so that $k_1+k_2+k_3=q_1+q_2$.
Note an important

Summation over  the colors of the final pomerons in the expression for the
amplitude (1) transforms
$\Gamma_{a_1a_2a_3a_4}(k_1,k_2,k_3,k_4|q_1,q_2)$ into
\[
\Gamma(k_1,k_2,k_3,k_4|q_1,q_2)=
\left(\frac{1}{N_c^2-1}\right)^2
\sum_{a_1,a_3}\Gamma_{a_1a_1a_3a_3}(k_1,k_2,k_3,k_4|q_1,q_2)
\]\beq=\frac{1}{2}q_1^2q^2_2
\Big[V(1,2,3,4)+\frac{1}{N_c^2-1}\Big(V(1,3,2,4)+V(1,4,3,2)\Big)\Big].
\eeq
In the high-colour limit only the first term remains.
The function $F(j_i,t_i)$ in the integrand for the amplitude (1)
becomes
\beq
F(j_i,t_i)=\int \frac{d^2q_1}{(2\pi)^2}
\frac{d^2k_1}{(2\pi)^2}\frac{d^2k_3}{(2\pi)^2}
\Gamma(k_1,k_2,k_3,k_4|q_1,q_2)
P_1(j_1;k_1,k_2)P_2(j_2;k_3,k_4)P(j_0;q_1,q_2)
\eeq
Note that the resulting vertex $\Gamma$ is no more symmetric in all
the gluons, which is a consequence of an unsymmetrical projection onto
the colour space of the final reggeons.
Putting in (1) $s_1=s_2=s$, passing  to rapidities
defined as
\beq
y=\ln\frac{s}{M^2},\ \ Y=\ln\frac{s}{s_0}
\eeq
and substituting the signature factors by their lowest
order values we  obtain for the amplitude the standard expression
\[
T_{3\to 3}=-is\frac{1}{2\pi}
\int \frac{d^2q_1}{(2\pi)^2}
\frac{d^2k_1}{(2\pi)^2}\frac{d^2k_3}{(2\pi)^2}\]\beq
\Gamma(k_1,k_2,k_3,k_4|q_1,q_2)
P_1(y;k_1,k_2)P_2(y;k_3,k_4)P(Y-y;q_1,q_2),
\eeq
where the pomerons in the $y$-representation are defined as
\beq
P(y)=\int \frac{dj}{2\pi i}e^{y(j-1)}P(j).
\eeq

This amplitude refers to the case when we have a simple triple
pomeron diagram corresponding to scattering on two centers. A more
important case is  scattering on many centers
(heavy nucleus) described by a sum of all fan
diagrams made of the BFKL pomerons with their triple interaction.
Apart from the driving term, which is an exchange of a single pomeron,
it is given by the expression of the same form
as (19) in which the final pomerons are
substituted by the sums of all fans $\Phi(y,k_1,k_2;b)$
and $\Phi(y,k_3,k_4;b)$ where
$b$ is the impact parameter.

Expression (19) can be further simplified taking the limit $N_s\to\infty$
and passing to the coordinate representation \cite{BV}.
Here we want only to comment about  different parts of the imaginary part of
this amplitude corresponding to different cuts.
They will generally involve different projections onto the color space of
the final pomerons. However due to symmetry of the vertex $\Gamma$
the final result will be the same. For instance in the double cut
we find the vertex
\beq
\left(\frac{1}{N_c^2-1}\right)^2
\sum_{a_1,a_3}\Gamma_{a_1a_3a_1a_3}(k_1,k_3,k_2,k_4|q_1,q_2).
\eeq
However the symmetry allows to interchange gluons 2 and 3 and (21) becomes
identical to (16).
So all contributions to the imaginary part will contain the same
vertex $\Gamma(k_1,k_2,k_3,k_4|q_1,q_2)$ defined by (16) and so will be
given
by the same expression (19) with factor $-i$ changed to 1, 2 and -4 for
the diffractive, double cut and single cut contributions respectively.

\section{Single inclusive cross-section}
To derive the single inclusive cross-section for emission of a gluon jet
at rapidity $y$ and of the transverse momentum $k$ we have to find it in the
intermediate states in the unitarity relation fror $T_{3\to 3}$. Inspection
of the diagrams in Fig. 3 shows that first one can extract the jet
from within the pomerons joined at the vertex.
Emission from the pomeron corresponds to "opening" the BFKL chain,
which is described in the coordinate space
by inserting the emission operator  ~\cite{BT}
\beq
V_k(r)=\frac{4\alpha_sN_c}{k^2} \stackrel{\leftarrow}\Delta e^{ikr}
\stackrel{\rightarrow}\Delta,
\eeq
that is  substituting the BFKL Green function at rapidity interval $Y$ by
\beq
G(Y;r_1,r_2)\to \int d^2r G(Y-y;r_1,r)V_k(r)G(y:r,r_2).
\eeq

In this way we find the single inclusive cross-section at fixed
impact parameter $b$ on two centers
corresponding
to the emission from the upper pomeron in $T_{3\to 3}$ (see Fig. 4$a$)
\beq
J^{(P)}(y,k)\equiv\frac{(2\pi)^3d\sigma^{(P)}}{dy d^2k d^2b}
=2\int d^2rP(Y-y;r)V_k(r)\Phi^{(2)}(y;r|b).
\eeq
Here $P(Y-y;r)$ is the initial pomeron in the coordinate representation.
Function $\Phi^{(2)}(y;r|b)$ is the contribution from the double interaction
with the nucleus due to the 3-pomeron vertex. In the momentum representation
\beq
\Phi^{(2)}(y;q_1,q_2|b)=T^2(b)\int
\frac{d^2k_1}{(2\pi)^2}\frac{d^2k_3}{(2\pi)^2}
P_1(y;k_1,k_2)P_2(y;k_3,k_4)\Gamma(k_1,k_2,k_3,k_4|q_1,q_2)
\eeq
with $q_1+q_2=0$.
The AGK relations (6)-(8) tell us that emissions from lower pomerons in
$\Phi$  (Fig. 4$b$) do
not give any contribution.
\begin{figure}
\hspace*{1cm}
\epsfig{file=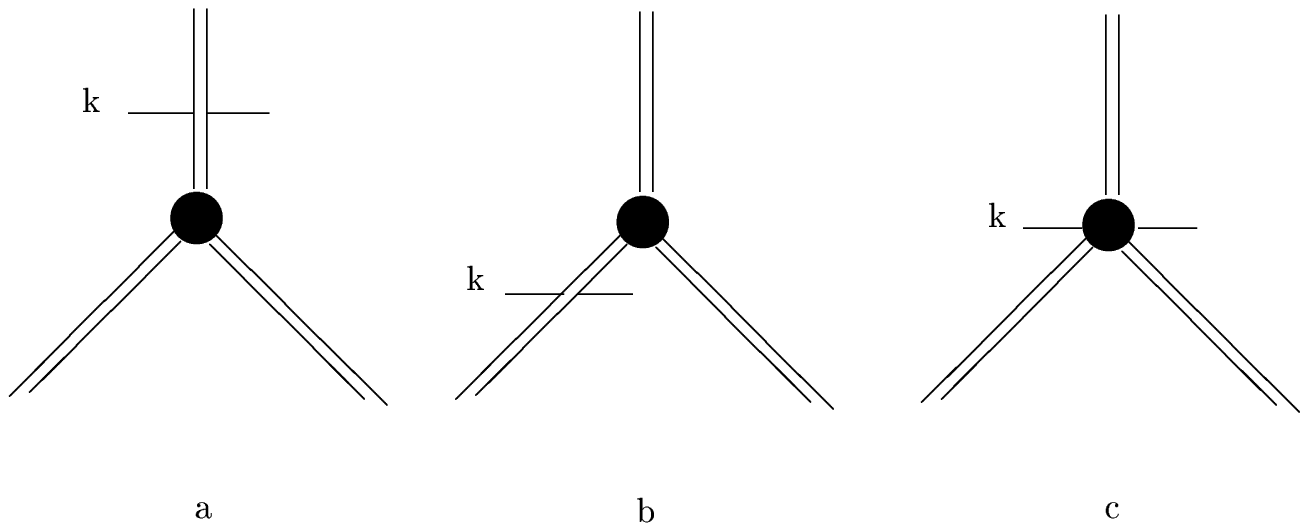,bbllx=119pt,bblly=559pt,bburx=536,bbury=740pt,width=10cm}
\caption{Pomeron diagrams for the single inclusive cross-section
on two centers}
\label{fig4}
\end{figure}

Thus we are left with the emission from within the vertex $\Gamma$
corresponding to the diagram shown in Fig. 4$c$.
To find this contribution it is evidently enough to study the
inclusive cross-section for the case when the three pomerons joined
at the vertex are taken in  the lowest order, the double gluon
exchange. In this case there is no gluon emission from inside the
pomerons and all gluons come either from the vertex or from the
additional contribution separated from the 3-pomeron diagram as the
mentioned reducible part in the reggeon diagram technique
or equivalently in the form of the Glauber rescattering in the
initial state in the evolution equation for the sum of fans $\Phi$.
All we have to do is to study in the lowest order all the
corresponding diagrams
with 4 gluon legs combined into the final pomerons
and locate the observed intermediate gluon
in the appropriate cuts. The number of initial gluons,
coupled to the quark-antiquark  loop may be 2,3 or 4. Correspondingly the
diagrams split into three types with transitions $2\to 4$, $3\to 4$ and
$4\to 4$ gluons.

Typical diagrams for these 3 cases are shown
in Figs. 5$a$-$c$.
\begin{figure}
\hspace*{2 cm}
\epsfig{file=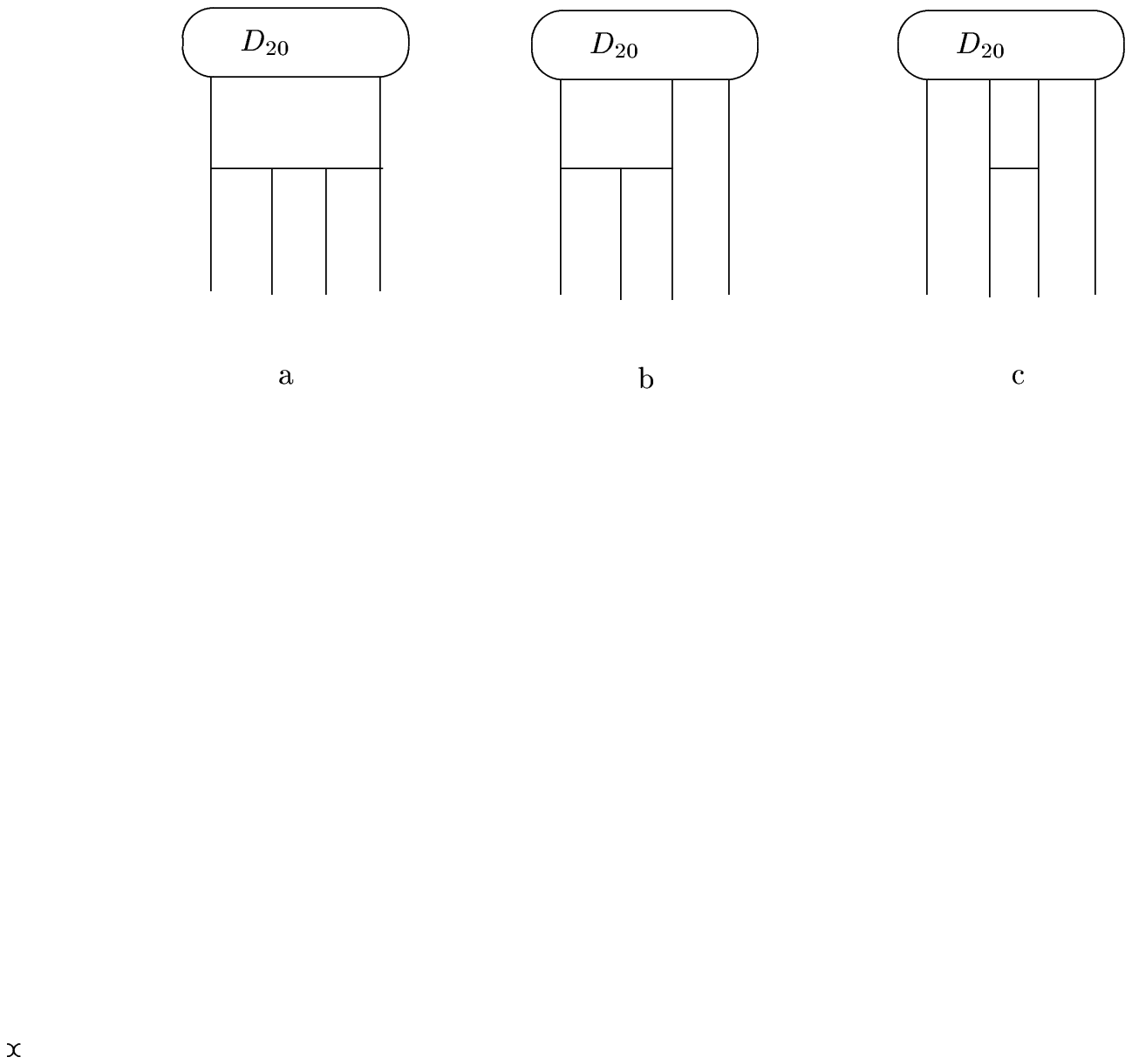,bbllx=155pt,bblly=571pt,bburx=531,bbury=727pt,width=10cm}
\caption{Transitions from 2 to 4 ($a$), 3 to 4 ($b$) and 4 to 4 ($c$) gluons}
\label{fig5}
\end{figure}
 The upper blob $D_{20}$ represents the quark-antiquark loop with 2, 3
or 4 gluons attached to it in all possible ways conserving the order of
gluons 1, 2, 3 and 4. The final gluons are to be understood as
parts of the lower pomerons in $T_{3\to 3}$. They are to be
combined into pomerons in two different configurations: the diffractive
one, in which the pomerons are made of pairs (1,2) and (34), and the
double cut one with pomerons made of pairs (1,4) and (2,3). Note that pairing
(1,3) and (2,4) is prohibited. It is well-known that two ladders can couple
to a particle only in the so-called "nested" configuration. This can also
be understood from the kinematical situation: the rungs in the pomeron ladders
do not depend on the longitudinal variables and so are instantaneous in the
longidudinal directions. Diffractive and double cut configuration allow the
relative longitudinal distance between the two pomerons to be arbitrary, which
translates into the integration of the  nuclear density into the profile
function. In the configuration (1,3) (2,4) the two pomerons are in fact
located  at the same longitudinal coordinate, which leads to suppression
of this contribution by the inverse of the large nuclear dimension.
The diffractive and double cuts are  made between the gluons
2 and 3 (see Figs. 3$a$ and $b$) . The single cuts are made in both
configurations separating gluons
1 or 4 (Figs. 3$c$ and $d$).
Actual calculations are best performed in the coordinate space (although
naturally they are completely equivalent to the calculations in the
momentum space  customary in the reggeon diagram technique).
The  vertexes for the transitions  $n\to 4$ gluons with $n=2,3,4$
can be conveniently represented via the emission vertexes reggeon$\to$
reggeon+particle (Lipatov) and reggeon$\to m$ reggeons+particle (Bartels)
which may be found in ~\cite{BW}. In the coordinate space emission
of a gluon of momentum $k$ at point
${\bf z}_1$ from the quark  at point ${\bf r}_1$
by the Lipatov vertex is described by factor
\beq
\fa=\ka-\ha,\ \ \ka=\frac{\bf k}{k^2}\delta^2(z_1-r_1),\ \
\ha=\frac{i}{2\pi}\frac{{\bf z}_1-{\bf r}_1}{(z_1-r_1)^2}.
\eeq
Similarly $\fb$ refers to the emission from antiquark with coordinate
${\bf r}_2$
And $\fc$ and $\fd$ refer to the conjugate amplitude with the emission
point ${\bf z}_2$.
Emission by the Bartels vertex  is described by factor
\beq
B_1(r_1,x)=\ha\Big(\delta^2(x-r_1)-\delta^2(x-z_1)\Big),
\eeq
where $\bf x$ is the coordinate of splitting of the first final gluon.
The rest of the final gluons are to be located at ${\bf z}_1$

Following ~\cite{KT} we present the result for this lowest order inclusive
cross-section in the form of the integral over the interquark distance
in the loop
$r_{12}=r_1-r_2$ and gluon emission points in the  direct and conjugate
amplitudes
\beq
J_0(y,k)\equiv\frac{(2\pi)^3d\sigma_0}{dy d^2k d^2b}
=g^2N_cT^2(b)\int
d^2r_{12}d^2z_1d^2z_2e^{-ikz}I(r_1,r_2,z_1,z_2)
\nabla^4P^{(0)}(r_{12}).
\eeq
Here $P^{(0)}(r)$ is the upper pomeron in the lowest order. Application of
$\nabla^4$ removes its legs and leaves only the quark-antiquark loop
in the coordinate representation $D_{20}(r)$.
The integrand $I$ is a sum of contributions from the diffractive,
double and single cuts:
\beq
I=I^{diff}+I^{double}+I^{single}.
\eeq
Each of them is just the sum of the contributions from the appropriately
cut diagrams with the statistical weight factors 2, 4 and -4 for the diffractive,
double and single cuts respectively.

To write down the contributions we use the
shorthand notations for the lower pomerons (we consider them identical)
\[
X\equiv P({\bf z}_1-{\bf r}_1),\ \ Y\equiv P({\bf z}_1-{\bf r}_2),\ \
U\equiv P({\bf z}_2-{\bf r}_1),\]\beq
V\equiv P({\bf z}_2-{\bf r}_2),\ \ Z\equiv P({\bf z}_1-{\bf z}_2),
\ \ R\equiv P({\bf r}_1-{\bf r}_2).
\eeq
As mentioned the pomerons are to be taken in the lowest order (two-gluon
exchange) and so do not depend on rapidity.

Simple, although somewhat tedious calculations, give the following results for
the contributions from the diffractive, double and single cuts.
\beq
I^{dif}=(\ha-\hb)(\hc-\hd)(X+Y-R)(U+V-R).
\eeq
\[
I^{double}=(\ha-\hb)(\hc-\hd)\Big[4Z^2
+2\Big(\pa\pd+\pb\pc\Big)
-3\pz\Big(\pa+\pb+\pc+\pd\Big)\Big]
\]\[
+\ha\hc\Big[U(R+3V-U)+X(R+3Y-X)\Big]\]\[
+\hb\hd\Big[V(R+3U-V)+Y(R+3X-Y)\Big]\]\[
-\ha\hd\Big[U(R+3V-U)+Y(R+3X-Y)\Big]\]\beq
-\hb\hc\Big[V(R+3U-V)+X(R+3Y-X)\Big].
\eeq
\[I^{single}=(\ha-\hb)(\hc-\hd)\Big\{3\pz
\Big[\pa+\pb+\pc+\pd\Big]
+\pa\pc+\pb\pd+2\pa\pd+2\pb\pc\Big\}
\]
\[
+(\ha U-\hb V)(\hc-\hb)\Big[-2R-3(U+V)\Big]\]\[
+(\ha-\hb)(\hc X-\hd Y)\Big[-2R-3(X+Y)\Big]\]\[
+\ha\hc R\Big(2V+3U+2Y+3X\Big)
+\hc\hd R\Big(2U+3V+2X+3Y\Big)\]\[
-\ha\hd R\Big(2V+3U+2X+3Y\Big)
-\ha\hb R\Big(2U+3V+2Y+3X\Big)\]\beq
-2R^2\Big(\ha+\hb\Big)\Big(\hc+\hd\Big).
\eeq

Remarkably in the sum of these three contributions nearly all
the terms cancel and one gets a comparatively simple expression
\[
I=4\ha\hc\Big(Z^2-X^2-U^2\Big)
+4\hb\hd\Big(Z^2-Y^2-V^2\Big)\]\beq
-4\ha\hd\Big(R^2+Z^2-Y^2-U^2\Big)
-4\hb\hc\Big(R^2+Z^2-X^2-V^2\Big).
\eeq
It agrees with the expression found
in the Glauber approach in ~\cite{KT}

As mentioned to find the contribution from the vertex one has to
subtract from this expression the term which comes from the reducible
part. In the coordinate representation this part comes from the
contribution of the double scattering in the Glauber expression for the
initial state:
\beq
T^{red}=\int d^2rP_1^{(0)}(r)P_2^{(0)}(r)P(Y,r),
\eeq
where the upper pomeron is a developed one, but the two lower
ones are to be taken in the lowest order. The contribution to the
inclusive cross-section is obtained by opening the upper pomeron
(and multiplying the result by 2 for the twice imaginary part).
The found inclusive cross-section can again be represented in the form
(28) with an integrand
\beq
I^{red}=\frac{1}{2}I.
\eeq
This brings us to the final contribution from the vertex, which has the form
(28) with an integrand
\beq
I^{vertex}=\frac{1}{2}I.
\eeq
The simple form of $I$ allows to do the integrations over $z_1$ and $z_2$
and present the inclusive cross-section from the vertex in a simpler form
similar to (28) (see ~\cite{KT}):
\beq
J^{(V)}_0(y,k)
=-T^2(b)\int d^2rP^{(0)}(r)V_k(r)[P^{(0)}(r)]^2.
\eeq

This contribution has been found for the case when all pomerons are taken
in the lowest order, without evlution. However from the structure of the
triple pomeron diagram it immediately follows that evolution just
restores all orders for all the three  pomerons, so that the inclusive
cross-section coming from emission from the vertex is given by the same
expression (38) with the  pomerons taken fully evolved,
the upper one up to $Y-y$ and the two lower ones up to $y$.
\beq
J^{(V)}(y,k)\equiv\frac{(2\pi)^3d\sigma^{(V)}}{dy d^2k d^2b}
=-\int d^2rP(Y-y;r)V_k(r)[\Phi^{(1)}(y;r|b)]^2.
\eeq
where $\Phi^{(1)}$ corresponds to a single interaction with the nucleus:
\beq
\Phi^{(1)}(y;r|b)=P(y;r)T(b).
\eeq

Passing to scattering on many centers we have to take into account
that due to the AGK cancellations ony the  contributions from the
uppermost pomeron and vertex remain. So to obtain the inclusive
cross-section we have only to appropriately
change the lower legs in the contributions from the single and double
scattering. In this way we get the final inclusive cross-section as
\beq
J(y,k)\equiv\frac{(2\pi)^3d\sigma}{dy d^2k d^2b}
=\int d^2r_1d^2rP(Y-y;r)V_k(r)\Big(2\Phi(y;r|b)-\Phi^2(y;r|b)\Big).
\eeq
The second term corresponds to emission from the vertex and agrees with
the result of ~\cite{KT}.

Note finally that integrating the found contributions from different cuts
over
$k$ we obtain the total imaginary parts of the amplitude due to emission of
a gluon from the
vertex, divided by $s$  and so coinciding
with the corresponding cross-sections $\sigma_1$:
\beq
\sigma_1^{dif}=4\alpha_s(\ha-\hb)^2(X+Y-R)^2,
\eeq
\[
\sigma _1^{double}=2\alpha_s\Big\{2h_1^2(r_1)(XR+5XY-X^2)+
2h_1^2(r_2)(YR+5XY-Y^2)\]\beq
-2h_1(r_1)h_1(r_2)\Big(R(X+Y)+10XY-X^2-Y^2\Big)\Big\},
\eeq
\[
\sigma_1^{single}=2\alpha_s\Big\{
2h_1^2(r_1)\Big(R(X+2Y)-7XY-4X^2-Y^2\Big)+
2h_1^2(r_2)\Big(R(2X+Y)-7XY-X^2-4Y^2\Big)\]\beq+
2h_1(r_1)h_2(r_2)\Big(3R(X+Y)+14XY+5X^2+5Y^2\Big)\Big\}.
\eeq
One observes that they do not satisfy the AGK relations (6)-(8).
Only summed with the cross-sections without emission of a gluon from
the vertex they do satisfy these relations.

\section{Evolution equation for the inclusive cross-section on two
centers}
Evolution of the inclusive cross-section with rapidities follows
directly from the representation (28) and evolution equations for
$P(Y-y;r)$ and $\Phi(y;r|b)$. At a fixed number of centers however
the structure of the inclusive cross-section allows to construct an
evolution equation directly for it, as an equation which describes
evolution of the lower legs. We shall limit ourselves with the
inclusive cross-section on two centers (two lower pomerons).

As mentioned, the amplitude $T_{3\to 3}$ itself can be split  into a
reducible and irreducible parts. Separating from the irreducible
part the nuclear sources with final  gluon propagators
one obtains the irreducible 4-gluon amplitude  $D_4^{(I)}$ studied in
~\cite{BW,BE} It satisfies
an equation  obvious from the representation (1), which can be written
symbolically as
\beq
\Big(\frac{\partial}{\partial y}+H_4\Big)D_4^{(I)}(y)=\Gamma P(y).
\eeq
Here $H_4$ is the Hamiltonian for 4 reggeized gluons which form the
final pomerons. $P(y)$ is the upper Pomeron.
$\Gamma$ is the 3-pomeron vertex as an operator acting from the space of
the two initial gluons into the space of the four final gluons.
In the limit $N_c\to\infty$ the Hamiltonian $H_4$ contains only the
BFKL interactions between the gluons inside each of the two pomerons.
The solution of Eq. (45) can be achieved by applying to the righthand side
the Green function for 4 reggeized gluons:
\beq
D^{(I)}_4(y)=\int dy'G_4(y-y')\Gamma P(y').
\eeq
In the limit $N_c\to \infty$ $G_4$ is just a product of two independent
BFKL Green functions for the two pomerons. Attaching the sources to (46)
one restores represetation (1).

Passing to the inclusive cross-section (41), for two scattering centers
we can introduce a similar
amplitude $D_4^{(I)}(Y,y|k)$, which is obtained by separating from the cross-section
the two nuclear sources and  changing the direction
of the evolution, so that the nucleus is at rapidity Y, the
splitting occurs at $y$ and the projectile is at zero rapidity.
From (41) we find
\beq
D_4^{(I)}(Y,y|k)=\int dy'G_4(Y-y')\Gamma P(y',y|k)-G_4(Y-y)V_kP(y),
\eeq
where $P(Y,y|k)$ is an opened pomeron:
\beq
P(Y,y|k)=G(Y-y)V_kP(y).
\eeq
Note that here $V_k$ acts exclusively in the 2-gluon space whereas in
(47) it acts from the 2-gluon into 4 gluon space. Obviously
$D_4^{(I)}(Y,y|k)$ satisfies an equation:
\beq
\Big(\frac{\partial}{\partial Y}+H_4\Big)D_4^{(I)}(Y,y|k))=
\Gamma P(Y,y|k)-\delta(Y-y)V_kP(y).
\eeq
Indeed applying the Green gunction $G_4$ to the right-hand side one
obtains (47).
Comparing (49) with (45) one observes that the equations are actually
quite similar
at $Y\ne y$. The role of the second term describing emisssion from the vertex
is only to supply the initial condition for the evolution at $Y=y$:
\beq
D^{(I)}_4(y,y|k))=-G(0)V_kP(y).
\eeq

\section{Double inclusive cross-section}
Now we study the double inclusive cross-section for emission of two jets
with rapidities and transverse momenta $y_1,k$ and $y_2,l$. We assume
$y_1>y_2$ (and in fact in the BFKL kinematics $y_1>>y_2$).
Both gluons may come from within the pomerons.
The AGK rules tell us that the only contributions of this sort
which remain are from either two emissions from the upper pomeron or
from emissions from the both lower pomerons immediately after the first
splitting shown in Fig.6$a$ and $b$. These are the standard AGK contributions.
\begin{figure}
\hspace*{2 cm}
\epsfig{file=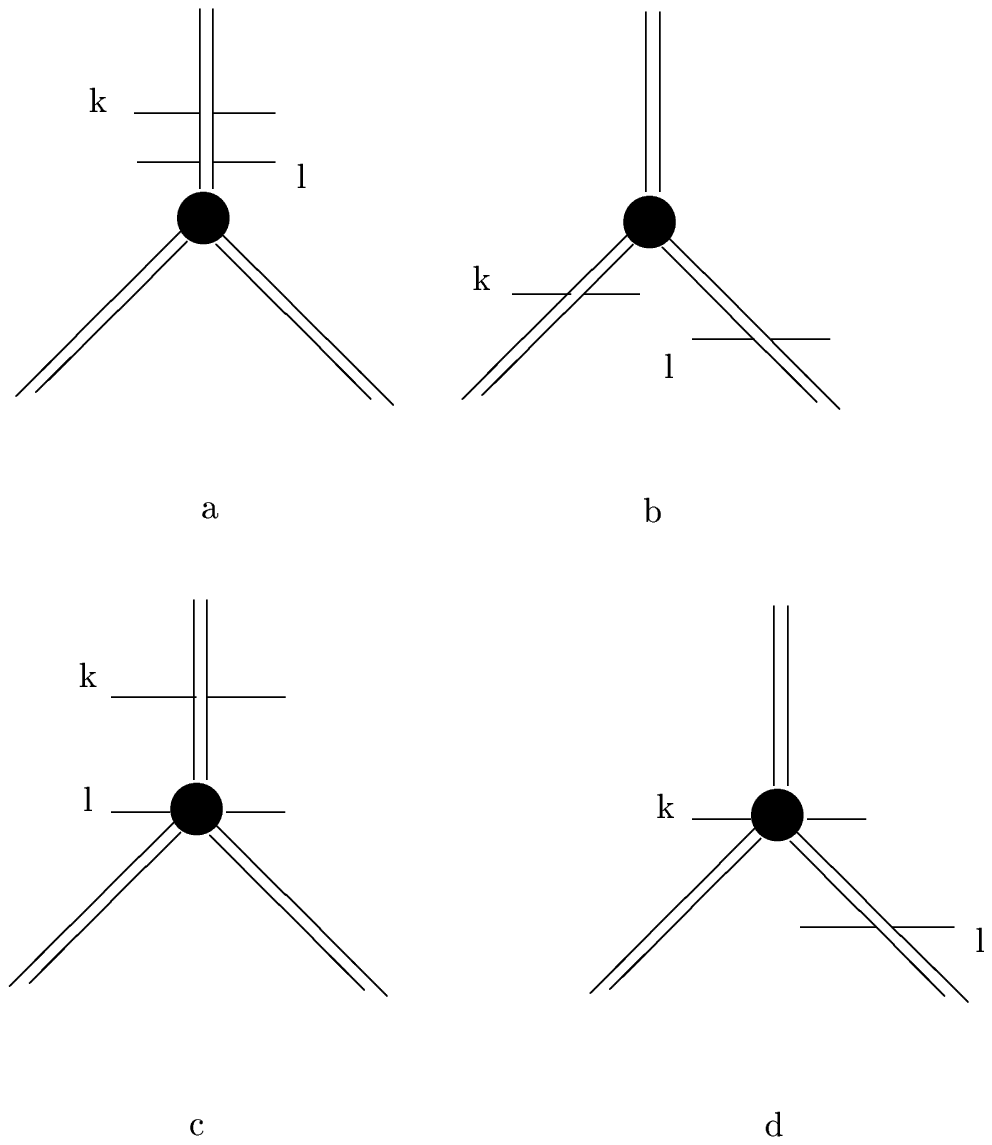,bbllx=178pt,bblly=366pt,bburx=513,bbury=717pt,width=8cm}
\caption{Pomeron diagrams for the double inclusive cross-section
on two centers}
\label{fig6}
\end{figure}

Now we pass to contributions which involve emission from the vertex.
Clearly in the lowest order there cannot occur two emissions from
the vertex at the same rapidity. According to the AGK
rules we are left with two cases:
either the faster gluon (rapidity $y_1$) is emitted from the uppermost
pomeron
and the slower one from the first splitting vertex (Fig. 6$c$)
or the faster gluon is
emitted from the uppermost
vertex and the slower one from one of the lower pomerons immediately
after the first splitting (Fig. 6$d$).

The first case is simple. In this case the lower pomerons can be
cut in any way and the contribution from the vertex includes all three
terms corresponding to diffractive, double and single cuts, which combine
into the final expression (38) for emission from the vertex.
So the double inclusive cross-section in this case is obtained from (38)
by just additionally 'opening' the upper pomeron:
\[
J^{(1)}(y_1,k;y_2,l)\equiv\frac{(2\pi)^6d\sigma}{dy_1 d^2k dy_2d^2l d^2b}
\]\beq
=-\int d^2rd^2r'P(Y-y_1;r)V_k(r)G(y_1-y_2;r,r')V_l(r')
\Phi^2(y_2;r'|b).
\eeq

The second case is more complicated. Now the the lower pomeron which
emits the gluon  must be cut. This excludes the diffractive part of
the contribution to the emission from the vertex and instead of the simple
expression (38) we have to use the sum of only double and single cuts
for the vertex emission.
The integrand in (28) results more complicated
\beq
I_1=(1/2)(I^{double}+I^{single})=(1/2)(I-I^{dif}).
\eeq
By shifting variables $z_1$ and $z_2$ one can present the contribution
to emission from the vertex from only the diffractive cut in the
form
\beq
J^{dif}(y,k)=T^2(b)\int d^2rd^2r_1d^2r_2 P_1(y,r_1)P_2(y,r_2)
P(Y-y,r)\Gamma^{dif}_k(r_1,r_2|r).
\eeq
The emission vertex $\Gamma^{dif}_k(r_1,r_2|r)$ can be presented via the
kernel $K$, Eq. (15), in which the momenta a substituted by coordinates
(not the Fourier transform!):
\[
\Gamma^{dif}_k(r_1,r_2|r)=g^2N_ce^{-ik(r_1-r_2)}\Big\{
K(r,r_1+r_2-2r,r|r_1,r_2)-e^{-ikr}K(r,r_1+r_2,-r|r_1,r_2)
\]\[
-\delta^2(r-r_1)
\int d^2r_2'K(r,r_1+r'_2-2r,r|r_1,r'_2)
-\delta^2(r-r_2)
\int d^2r_1'K(r,r'_1,r_2-2r,r|r'_1,r_2)\]\beq
+\frac{1}{2}\delta^2(r-r_1)\delta^2(r-r_2)
\int d^2r'_1d^2r_2'
K(r,r'_1+r'_2-2r,r|r'_1,r'_2)\Big\}.
\eeq
The double inclusive cross-section with emission of the fastest
gluon jet from the vertex and the other from the lower pomeron will be given
by 'opening' in (38) the lower pomeron and subtracting from the contribution
from the vertex the diffractive cut part:
\[
J^{(2)}(y_1,k;y_2,l)
=-\int d^2rP(Y-y_1,r)V_k(r)\Phi(y;r|b)
G(y_1-y_2;r,r')V_l(r')\Phi(y_2;r'|b)\]\[-
\int d^2r d^2r_1 d^2r_1'd^2r_2P(Y-y_1,r)\Gamma_k^{dif}(r_1,r_2,r)
\Phi(y;r_2|b)G(y_1-y_2;r_1,r'_1)V_l(r'_1)\Phi(y_2;r'_1|b)\]\beq
+(\rr).
\eeq

\section{Conclusions}
We have established that the AGK rules can only be satisfied if the triple
pomeron vertex is a fully symmetric function in all four reggeized gluons
which  form the two outgoing pomerons. This selects the symmetric Bartels
vertex $V$ as a vertex for the triple pomeron amplitude obeying the
AGK rules. The total amplitude for the scattering on two centers thus
splits into this triple pomeron part and a single pomeron exchange one, both
satisfying the AGK rules in their own way (the latter trivially).
The unitarity content for the triple pomeron amplitude
allows to determine the contribution from gluon emissions from inside
the pomeron in a straightforward manner. The contribution from the vertex
emission can be derived neglecting the evolution of the pomerons  and summing
the relevant reggeized gluon diagrams. For the single inclusive cross-section
this leads to the result obtained  in ~\cite{KT} from the dipole picture.
   The 'opened' triple pomeron vertex describing gluon emisssion from it
is found to depend on the nature of the cut through it, which point was
mentioned when the AGK rules were derived in ~\cite{AGK}. As a result, in the
double inclusive cross-section with the slower gluon emitted after the
splitting the cut opened  vertex is different  and more complicated than in
the single inclusive cross-section, since the contribution from
the diffractive cut has to be dropped. Still the fundamental AGK properties
are found to be valid also for the double inclusive cross-section, so that
a contribution from the upper pomeron and the one after the first
branching claimed in  ~\cite{JK} does not appear.

As a byproduct we constructed an evolution equation for the single
inclusive cross-section on two centers as a function of the overall rapidity.
This equation may be helpful in generalizing to the case of finite $N_c$,
when the final pomerons interact between themselves.

\section{Acknowledgements}
The author benefited from numerous and highly constructive discussions
with J.Bartels, M.Salvadore and G.P.Vacca. He is grateful to the
Universities of Hamburg and Santiago de Compostela  for hospitality.
This work was supported  by grants of the
Education Ministry of Russia RNP 2.1.1.1112 and RFFI Russia 06-02-16115-a.

\end{document}